\documentstyle[amsfonts]{article}

\input epsf

\newcommand{\R}{{\mathbb R}}
\newcommand{\Z}{{\mathbb Z}}
\newcommand{\N}{{\mathbb N}}
\newcommand{\beq}{\begin{equation}}
\newcommand{\eeq}{\end{equation}}
\newcommand{\bea}{\begin{eqnarray}}
\newcommand{\eea}{\end{eqnarray}}
\newcommand{\ra}{\rightarrow}
\newcommand{\ds}{\displaystyle}
\newcommand{\cd}{\partial}
\newcommand{\B}{\overline{B}}

\newcommand{\wt}{\widetilde}
\newcommand{\wh}{\widehat}
\newcommand{\T}{{\rm T}}
\newcommand{\IM}{{\mathbb I}}

\newtheorem{thm}{Theorem}
\newtheorem{lemma}[thm]{Lemma}
\newtheorem{prop}[thm]{Proposition}
\newtheorem{cor}[thm]{Corollary}
\newtheorem{defn}[thm]{Definition}

\newcommand{\am}{{\rm am}}
\newcommand{\sn}{{\rm sn}}

\newcommand{\bPsi}{\mbox{\boldmath $\Psi$}}

\newcommand{\bTheta}{\mbox{\boldmath $\Theta$}}

\newcommand{\K}{{\rm ker}}
\newcommand{\ctwoloops}{\mbox{$\Omega_{T,2}^{-}$}}
\newcommand{\czeroloops}{\mbox{$\Omega_{T,0}^{-}$}}
\newcommand{\eqref}{\ref}
\newcommand{\z}{\bf 0}

\begin{document}

\title{Breathers in the weakly coupled
topological discrete sine-Gordon system}
\author{M. Haskins\thanks{E-mail: {\tt mhaskins@math.utexas.edu}}
\, and J.M. Speight\thanks{Present address: Max-Planck-Institut f\"{u}r
Mathematik in den Naturwissenschaften, Inselstra\ss e 22-26,
Leipzig, Germany. E-mail: {\tt speight@math.utexas.edu}}\\
Department of Mathematics, University of Texas at Austin\\
Austin, Texas 78712, U.S.A.}

\date{}

\maketitle
\begin{abstract}
Existence of breather (spatially localized, time periodic, oscillatory)
solutions of the topological discrete sine-Gordon (TDSG) system, in the
regime of weak coupling, is proved.
The novelty of this result is that, unlike the systems previously considered
in studies of discrete breathers, the TDSG system does not decouple into
independent oscillator units in the weak coupling limit. The results of a
systematic numerical study of these breathers are presented, including
breather initial profiles and a portrait of their domain of existence in the
frequency--coupling parameter space. It is found that the breathers are
uniformly qualitatively different from those found in conventional 
spatially discrete
systems.
\end{abstract}

\section{Introduction}
Soliton dynamics in spatially discrete nonlinear systems plays a crucial part in the modeling
of many phenomena in condensed matter and biophysics.  One also encounters such systems in the numerical
simulation of soliton dynamics in high energy physics where they arise as approximations of 
continuum models. In this case one is concerned with the extent to which these discretizations resemble 
their continuum counterparts -- of course one hopes for a close resemblance. On the other hand, for 
applications in condensed matter and biophysics it is crucial that in certain respects the discrete 
systems have features not found in their continuum counterparts. One aspect in which recent work
has revealed such a distinction is in the appearance of so-called breather solutions -- that is
spatially localized, time periodic solutions. Such solutions
are conventionally called breathers, by analogy with the breathers of the
continuum sine-Gordon equation. While it is rare for a nonlinear partial differential equation to
possess breather solutions (such solutions seem to be a feature of integrable
systems), there has been mounting evidence that they are ubiquitous 
in spatially discrete systems. After numerical work and some heuristic arguments for the
existence of such solutions, the existence of breathers in a weak coupling limit of a class 
of discrete systems was proved by MacKay and Aubry \cite{aubry-mackay}. Subsequently these ideas 
were extended to a wider class of discrete systems by Sepulchre and MacKay \cite{mackay-sepulchre}.
The idea of both proofs is that
each model has a parameter $\alpha$, the coupling constant, and in the special
case $\alpha=0$ (called variously the anti-continuum, anti-integrable or
weak coupling limit) the system decouples into independent oscillator units.
If one sets one such unit oscillating while holding all others at rest, one
obtains a trivial one-site (or unit) breather. Persistence of breathers 
in a sufficiently small neighbourhood of
$\alpha=0$ then follows from an implicit function theorem argument.

In this paper we present a study of breathers in the topological discrete
sine-Gordon (TDSG) system \cite{ward-speight}. 
The system is so called because it reduces to the 
sine-Gordon equation in the continuum limit, and because it preserves many
of the ``topological'' features of this equation, namely a topological lower 
bound on kink energy, and a continuous moduli space of static kinks saturating
this bound. As a consequence of the preservation of these properties, kink
dynamics in this system is remarkably close to that of its continuum
counterpart, even in the highly discrete regime (kinks experience no
Peierls-Nabarro barrier, and may propagate freely, indefinitely). One 
motivation for the present work is to see whether this continuum-like
behaviour extends to breathers also. The TDSG system falls just outside the scope of 
the existence theorems actually proved in \cite{aubry-mackay} and \cite{mackay-sepulchre} 
(although both papers indicate that their methods \emph{should} apply in situations 
not directly covered by the theorems they state).
The reason is that there is no limit in which the system decouples into independent units.
Nevertheless, the analysis of the model and, in particular, its kink solutions suggest that there
is an anti-continuum-like limit. For example, there are one-site breathers in this limit and one suspects
that despite the lack of complete decoupling the continuation methods of MacKay et al.\ can be adapted
to this setting. We show that this is indeed the case and prove a theorem about constant period continuation 
of the one-site breathers.

Furthermore, the continuation can be performed numerically by searching for fixed points
of the period mapping of the system using a Newton-Raphson method, a 
technique developed for other systems by Aubry and Mar\'{\i}n \cite{aubry-marin}. In 
this way we obtain a portrait of the period-coupling parameter space showing
how far breathers of any particular period can be continued away from
the zero coupling limit. We also obtain
approximate breather initial profiles, and these display a rather unusual
feature. Namely, the sites adjacent to the central oscillating site
(the only one moving in the $\alpha=0$ limit) shift away from equilibrium
in the opposite direction to the central site. So the initial profiles are
all, roughly speaking, sombrero shaped, rather than hump shaped as have 
usually been observed in discrete systems. The direction of continuation of 
breathers away from $\alpha=0$ can be computed by a completely different
(and very simple) numerical method, providing independent confirmation of this
behaviour. The conclusion is, then, that breathers in the extremely discrete
regime of the TDSG system are qualitatively quite different from those which
appear in its continuum limit, and the two types of breather appear to be
unrelated.

The rest of this paper is arranged as follows. Section 2 introduces the
TDSG system and derives the one-site breathers in its weak coupling limit.
In section 3, persistence of these breathers away from zero coupling, and
exponential spatial localization of the continued breathers are proved.
Section 4 presents the results of the numerical search for breathers using
the Newton-Raphson method. In section 5 the direction of continuation is 
rederived by independent means. The work is summarized in section 6.

\section{The topological discrete sine-Gordon system}
\label{sec:TDSG}
The TDSG system (introduced by Ward and first studied in \cite{ward-speight})
is a field theory in $1+1$ dimensions where space is discrete,
with lattice spacing $h$, and time is continuous. It is most conveniently
defined in the Lagrangian formalism. To each two-sided sequence $\psi:\Z\ra\R$
one associates the potential energy
\begin{equation}
E_{P}[\psi]=\frac{h}{4}\sum_{n\in\Z}(D_{n}^{2}+F_{n}^{2})
\end{equation}
where
\begin{equation}
D_{n}=\frac{2}{h}\sin\frac{1}{2}(\psi_{n+1}-\psi_{n}),\qquad
F_{n}=\sin\frac{1}{2}(\psi_{n+1}+\psi_{n}).
\end{equation}
In the limit $h\ra 0$, $D\ra\psi_{x}$ and $F\ra\sin\psi$, so $E_{P}$ reduces to
\begin{equation}
E_{P}^{cont}[\psi]=
\frac{1}{4}\int_{-\infty}^{\infty}(\psi_{x}^{2}+\sin^{2}\psi)dx,
\end{equation}
the potential energy functional of the continuum sine-Gordon model. The key
feature of this particular discretization is that
\begin{equation}
D_{n}F_{n}=-\frac{1}{h}(\cos\psi_{n+1}-\cos\psi_{n}),
\end{equation}
so any configuration $\psi$ satisfying kink boundary conditions
\begin{equation}
\label{eq:kinkbc}
\ds{\lim_{n\ra-\infty}\psi_{n}=0},\qquad\ds{\lim_{n\ra\infty}\psi_{n}=\pi}
\end{equation}
 is subject
to the inequality
\begin{equation}
0\leq\frac{h}{4}\sum_{n\in\Z}(D_{n}-F_{n})^{2}=
E_{P}+\frac{1}{2}\sum_{n\in\Z}(\cos\psi_{n+1}-\cos\psi_{n})
=E_{P}-1.
\end{equation}
So $E_{P}\geq 1$ with equality if and only if $D_{n}=F_{n}$ for all $n$.
Solutions of this first order difference equation for $\psi_{n}$ are minimizers
of $E_{P}$ among all sequences satisfying the kink boundary conditions (\ref{eq:kinkbc}), and hence static solutions of the model.
The general solution of the first order difference equation is
\begin{equation}
\psi_{n}=2\tan^{-1}e^{a(nh-b)}
\label{eq:kink}
\end{equation}
where $a=h^{-1}\log[(2+h)/(2-h)]$ and $b$ is a real valued parameter, 
identified as the kink position. So, rather remarkably, the TDSG system has
a continuous moduli space of static kinks (parametrized by $b\in\R$) and
no Peierls-Nabarro barrier ($E_{P}=1$ independent of $b$). 

Dynamics is introduced into the system by defining the obvious kinetic energy
\begin{equation}
E_{K}=\frac{h}{4}\sum_{n\in\Z}\dot{\psi}_{n}^{2}.
\end{equation}
The action functional is then $S[\psi]=\int(E_{K}-E_{P})dt$. The Euler-Lagrange equations
associated with this functional,
\begin{equation}
\label{eq:motion}
\ddot{\psi}_{n}=\frac{4-h^{2}}{4h^{2}}\cos\psi_{n}(\sin\psi_{n+1}+
\sin\psi_{n-1})-\frac{4+h^{2}}{4h^{2}}\sin\psi_{n}(\cos\psi_{n+1}+
\cos\psi_{n-1})
\end{equation}
provide the equations of motion for the TDSG.
The analogue of the anti-continuum limit of this discrete sytem is not $h=\infty$, but rather
$h=2$, where the coefficient in front of the first term of (\ref{eq:motion}) vanishes,
and where the static kink solutions (\ref{eq:kink}) degenerate to step-like
functions. It is convenient to define a coupling constant 
\begin{equation}
\label{eq:alpha}
\alpha=\frac{4-h^{2}}{4h^{2}}
\end{equation}
so that the ``anti-continuum limit'' is $\alpha=0$. In this limit the equations of
motion reduce to
\begin{equation}
\label{eq:zerolim}
\ddot{\psi}_{n}=-\frac{1}{2}\sin\psi_{n}(\cos\psi_{n+1}+\cos\psi_{n-1}).
\end{equation}
Note that the lattice does not decouple into separate oscillators, but
maintains nearest-neighbour interactions. Nevertheless, as was observed in
\cite{ward-speight} the $\alpha=0$ system (\ref{eq:zerolim}) has one-site breather solutions
\begin{equation}
\label{eq:onesite}
\psi_{n}(t)=\left\{
\begin{array}{ll}
0 & n\neq 0 \\
\theta(t) & n=0
\end{array}\right.
\end{equation}
where $\theta(t)$ is any periodic solution of the pendulum equation
\begin{equation}
\label{eq:pendulum}
\ddot{\theta}+\sin\theta=0.
\end{equation}
We have chosen to locate the one-site breather at $n=0$, but clearly any other
site would do just as well. In fact, one could construct multisite breathers
where any number of sites, none adjacent to any other, oscillate according to
the pendulum equation, with periods in rational ratio. In this paper we will
consider only one-site breathers, however. 

As a system of autonomous differential equations, there is an
 $\mathbb{R}$-action on solutions of the 
TDSG system corresponding
to the time translation symmetry of the system. Thus any breather solution gives rise 
to a 1-parameter family of breathers. As a consequence, a breather solution is never an 
\emph{isolated} solution \cite{farkas}. To apply an implicit function theorem argument
one needs to deal with this degeneracy -- on the full phase space of the system the 
linearization of the equations of motion at a breather always has a non-trivial periodic solution
which generates time translation along this breather. We deal with this by
considering a subset of the phase space corresponding to solutions with a certain 
time reversal symmetry. We choose to work with solutions which are odd with respect to the time 
reversal symmetry i.e.\
 solutions that satisfy $\bPsi(-t)=-\bPsi(t)$ cf. \cite{aubry-mackay}.
The main reason for this choice of symmetry 
is simply that odd solutions of the pendulum equation are simpler to write
down in terms of elliptic functions than are even solutions. We will then prove
that the continuation theorem in the space of odd functions implies a similar
result in the space of even functions. Were we to consider continuation of
multisite breathers, we would need to remove several phase degeneracies
(one for each excited site) in order to apply the implicit function theorem. In
fact the same procedure (restriction to odd solutions) removes all such
degeneracies.

\section{Continuation of one-site breathers: analytic results}
\label{sec:analytic}
Our proof of the existence of discrete breathers for the TDSG system in the weak coupling regime follows the method used by Aubry and MacKay in \cite{aubry-mackay}. The proof proceeds in two steps; the first shows that certain periodic
solutions persist and the second that these solutions continue to be exponentially localized. In this section, after introducing the function spaces we choose to work with, we state and prove the two halves of the continuation theorem.

\subsection{The function spaces}
\label{sec:fnspaces}
Let $C^-_{T,2}$ denote the space of $C^2$ real-valued $T$-periodic functions 
on $\mathbb{R}$ with odd time reversal
symmetry. Equipped with the standard $C^2$ norm
\begin{equation}
  |\psi|_2 = \sup_{t\in \mathbb{R}}{\max{\{|\psi(t)|,|\dot{\psi}(t)|,|\ddot{\psi}(t)|\}}},
\label{eq:c2norm}
\end{equation}
$C^-_{T,2}$ becomes a Banach space.
Similarly, let $C^-_{T,0}$ denote the space of $C^0$ real-valued $T$-periodic functions
on $\mathbb{R}$ with odd time reversal symmetry, equipped with the standard $C^0$ norm
\begin{equation}
    |\psi|_0 = \sup_{t\in \mathbb{R}}{\{|\psi(t)|}\}.
\end{equation}
Consider the linear space $\oplus_{n\in \mathbb{Z}} X_n$, where each $X_n=C^-_{T,2}$, equipped with 
the natural sup norm i.e. if $\bPsi = (\psi_n)_{n\in \mathbb{Z}} \in\oplus_{n\in \mathbb{Z}} X_n$ 
then
\begin{equation}
{\|\bPsi\|}_2 = \sup_{n \in \mathbb{Z}}{{|\Psi_n|}_2}.
\end{equation}
\begin{defn}
$\ctwoloops := \{\bPsi\in \oplus_{n\in \Z}X_n : \|\bPsi\|_2<\infty\}$  
\end{defn}
\noindent
Similarly, equip the linear space $\oplus_{n\in \Z} Y_n$, 
where each $Y_n=C^-_{T,0}$, with its natural sup norm 
i.e. if $\bPsi = (\psi_n)_{n\in \Z} \in \oplus_{n\in \Z} Y_n$  then
\begin{equation}
{\|\bPsi\|}_0 = \sup_{n \in \Z}{{|\Psi_n|}_0}.
\label{eq:c0norm}
\end{equation}
\begin{defn}
  $\czeroloops := \{\bPsi\in \oplus_{n\in \Z}Y_n : \|\bPsi\|_0<\infty\}$  
\end{defn}
\noindent
Both $\ctwoloops$ and $\czeroloops$ are Banach spaces.

The equations of motion (\ref{eq:motion}) give a natural map 
$F: \ctwoloops\oplus\R\ra \czeroloops$ defined by
\begin{equation}
\label{eq:f}
  F(\bPsi,\alpha)_n = \ddot{\psi_n}-\alpha\cos{\psi_n}(\sin{\psi_{n+1}}+\sin{\psi_{n-1}}) + 
  (\alpha+\frac{1}{2})\sin{\psi_n}(\cos{\psi_{n+1}}+\cos{\psi_{n-1}}).
\end{equation}
The zeros of $F$ correspond precisely to the odd $T$-periodic solutions of the equations of motion
(\ref{eq:motion}).

In section \ref{sec:TDSG} we argued that $\alpha = 0$ should be considered as something akin
 to an anti-continuous limit for the TDSG system. Although the equations of 
motion~(\ref{eq:zerolim}) do not (in the terminology of MacKay and Sepulchre) decouple into 
\emph{local units}, the energy density of the kink solutions becomes concentrated at one site 
-- as one would demand for an anti-continuous limit. Moreover, like the models considered by 
Aubry and MacKay, at $\alpha = 0$ the system admits single site breather solutions -- 
in our case the single non-stationary site moves according to the pendulum equation.
So, analogously to the continuation of one-site breathers from the anti-continuous limit considered
by various authors \cite{aubry-mackay,mackay-sepulchre,aubry-marin}, 
one can ask
what happens to these solutions as the coupling $\alpha$ is increased. Despite the somewhat 
different nature of the TDSG system we find similar continuation results to those of the 
authors mentioned above. 
\subsection{Statement of the continuation theorem}

Given $T\in(2\pi,\infty)\setminus2\pi\mathbb{N}$, denote by $\theta$
 the unique odd $T$-periodic solution of the 
pendulum equation (\ref{eq:pendulum}) with positive initial velocity. Let $\bTheta$ denote the corresponding 
one-site breather solution of the TDSG at $\alpha=0$
given by (\ref{eq:onesite}). Then we have the following theorem.
\begin{thm}
\label{thm:cont}
There exists $\epsilon>0$ such that for $\alpha\in(0,\epsilon)$ there is a unique continuous family, $\bTheta_\alpha$, of odd $T$-periodic 
solutions of TDSG at coupling $\alpha$, such that $\bTheta_0=\bTheta$. Moreover, these 
solutions are exponentially localized, i.e.\ there exist positive constants
$B$ and $C$ (depending on $T$ and $\alpha$) such that \mbox{${|(\bTheta_\alpha)_n|}_0\le C\exp(-B|n|)$} holds for all $n\in\mathbb{Z}$.
\end{thm}
In other words, for small enough $\alpha$ the one-site breather $\bTheta$ has a locally unique continuation among odd $T$-periodic solutions
and these continued solutions remain exponentially localized.
The proof of this theorem is given in the next two sections. Section \ref{sec:periodproof} 
establishes the continuation through $T$-periodic solutions while section \ref{sec:local} 
deals with the question of localization.
\subsection{Persistence of periodicity}
\label{sec:periodproof}

By our remark in section \ref{sec:fnspaces}, $\bPsi$ is an odd solution of the TDSG system with period $T$ 
for coupling $\alpha$, if and only if, \mbox{$F(\bPsi, \alpha) = \z$} for 
the function $F$ defined by equations~(\ref{eq:f}). The odd one-site breather $\bTheta$ of period 
$T$ gives us a solution for $\alpha = 0$. The proof proceeds by showing that $F$ is a 
$C^1$ function to which the implicit function theorem may be applied. Thus a locally unique zero 
(and hence an odd $T$-periodic solution of TDSG) of $F$ for $\alpha$ close enough to 0 is demonstrated. 
It remains to show that the implicit function theorem can be applied to $F$,
 i.e.\ we need to check that 
$F$ is $C^1$ and that $DF$, its derivative with respect to the 
$\ctwoloops$\ factor is invertible at the one-site breather solution.

\vspace{0.5cm}
\noindent {\it Proof:}
The verification of the first fact is straightforward. A short computation gives us the following. 
If we let
\begin{equation}
{\bf D}F:(\ctwoloops\oplus\mathbb{R})\oplus(\ctwoloops\oplus\mathbb{R})\longrightarrow\czeroloops
\end{equation} be the derivative of $F$ and write 
\begin{equation}
{\bf D}F_{\Psi,\alpha}\,(\delta\bPsi,\delta\alpha) = (\widetilde{\psi}_n)_{n\in\mathbb{Z}}
\end{equation}
then ${\bf D}F$ is defined by
\bea
\widetilde{\psi}_n &=&
\ddot{\delta\psi_n}-\left[\sin{\psi_n}\left(\cos{\psi_{n+1}} + \cos{\psi_{n-1}}\right) - \cos{\psi_n}\left(\sin{\psi_{n+1}} + \sin{\psi_{n-1}}\right)\right] \delta \alpha \nonumber \\
& &+\left[\alpha\sin{\psi_n}\left(\sin{\psi_{n+1}} + \sin{\psi_{n-1}}\right)
+\beta\cos{\psi_n}\left(\cos{\psi_{n+1}} + \cos{\psi_{n-1}}\right)\right]\,\delta\psi_n \nonumber \\
& &- \alpha\cos{\psi_n}\left(\cos{\psi_{n+1}}\,\delta\psi_{n+1} + \cos{\psi_{n-1}}\,\delta\psi_{n-1} \right) \nonumber \\
\label{DERIV}
& &+\beta\,\sin{\psi_n}\left(\sin{\psi_{n+1}}\,\delta\psi_{n+1} + \,\sin{\psi_{n-1}}\,\delta\psi_{n-1}\right),
\eea
where $\beta=\alpha+\frac{1}{2}$. 
From this it is easily seen that ${\bf D}F$ is bounded.

At $\alpha=0$, $DF$, the derivative of $F$ with respect to the $\ctwoloops$ factor, reduces to
\bea
DF_{\Psi}(\delta\bPsi)_n &=& \delta\ddot{\psi}_n +
\frac{1}{2}\cos{\psi_n}(\cos{\psi_{n+1}} + \cos{\psi_{n-1}})\delta\psi_n 
\nonumber \\
\label{DERIV2}
& &
+ \frac{1}{2}\sin{\psi_n}(\sin{\psi_{n+1}}\delta\psi_{n+1} + \sin{\psi_{n-1}}\delta\psi_{n-1}).
\eea
In general this is a tridiagonal but not diagonal operator. However, $DF$
evaluated at any one-site breather is diagonal. Explicitly,
substituting $\bPsi=\bTheta$, the 
one-site breather of the form (\ref{eq:onesite}),
into (\ref{DERIV2}) yields
$DF_{\Theta} = \oplus_{n\in \mathbb{Z}} L_n$, where 
$L_n:C^-_{T,2}\ra C^-_{T,0}$ is,
\beq
L_n(\psi) =\left\{
  \begin{array}{ll}
    \ddot{\psi} + \cos{\theta}\,\psi & \mbox{if $n$ = 0},\\
    \ddot{\psi} + \frac{1}{2}(1 + \cos{\theta})\,\psi & 
    \mbox{if $n = \pm 1$},\\
    \ddot{\psi} + \psi& \mbox{otherwise}.
  \end{array}\right.
\eeq
Clearly, to prove invertibility of $DF$ at a one-site breather it is sufficient to prove that each
operator $L_n$ is invertible, since
then $DF^{-1}=\oplus_{n\in \mathbb{Z}} L^{-1}_n$ and
we see that $\|DF^{-1}\|\le \max{\left\{|L^{-1}_0|,|L^{-1}_1|,|L^{-1}_2|\right\}}$. Invertibility
of the operator $L_n$ will follow 
(see section \ref{sec:invert})
from standard facts about linear differential equations once
we establish that $\K{L_n}=\{0\}$.

\subsubsection{Injectivity of $DF_{\Theta}$} 

For $T\notin 2\pi\Z$, one sees immediately that $\K{L_n} =\{0\}$ for $|n|>1$.
So it remains to consider the central ($n=0$) and off-central ($n=\pm 1$)
equations
\bea
\ddot{\psi} + \cos{\theta}\,\psi&=&0 \label{C} \\
\ddot{\psi} + \frac{1}{2}(1 + \cos{\theta})\psi&=&0 
\label{OC} 
\eea
where $\theta$ is a $T$-periodic, odd solution of the pendulum equation,
$
\ddot{\theta} = -\sin{\theta},
$
with positive initial velocity. Both (\ref{C}) and (\ref{OC}) are Hill 
equations, that is ODE's of the form $\ddot{\psi}=q(t)\psi$ where $q$ is
periodic. Such equations have a long history. They arise, for example, in
stability analyses of celestial mechanics \cite{hill}. Note that in both
(\ref{C}) and (\ref{OC}) the potential, $q(t)$, has basic (i.e.\ smallest) period
$T/2$ since $\cos$ is even. Such equations may have periodic solutions only
with basic period not less than
$T/2$. To prove that $L_{0}$ and $L_{\pm 1}$
have trivial kernel, we will have to prove separately the non-existence of
$T/2$ and $T$-periodic odd solutions of (\ref{C}) and (\ref{OC}).

This task is simplified slightly for equation (\ref{C}) by the observation that
$\psi=\dot{\theta}$ is an {\em even\/} $T$-periodic solution. This is, in fact,
a consequence of the time translation symmetry of one-site breather solutions
previously mentioned. A standard result of Floquet theory \cite{Floquet}
 implies
that (\ref{C}) cannot have both  $T/2$-periodic and $T$-periodic solutions,
so it suffices in this case to rule out existence of $T$-periodic odd 
solutions. No such simplification occurs for the off-central equation
(\ref{OC}).

To proceed further one must consider the specific form of the solution
$\theta$ of the pendulum equation. If $\theta(0)=0$ and $\dot{\theta}(0)=
2k$, where $k\in(0,1)$ then the solution has amplitude of oscillation
$\theta_{0}\in(0,\pi)$ satisfying $\theta_0=2\arcsin{k}$, and period
$T=4K$ where
\beq
K=\int_{0}^{\frac{\pi}{2}}\frac{dz}{\sqrt{1-k^2\sin^2z}}
\eeq
is the complete elliptic integral of the first kind \cite{lawden}. 
The parameter $k$ is
called the modulus in the literature on elliptic functions. In fact this 
solution satisfies
\beq
\label{Jac}
\sin\frac{1}{2}\theta(t)=k\, \sn\, t
\eeq
where $\sn$ is the sn-oidal elliptic function of Jacobi. Note that $\sn$
itself, like all elliptic functions, depends parametrically on $k$. This
dependence will always be suppressed in our notation. Substitution of 
(\ref{Jac}) into (\ref{C},\ref{OC}) recasts them as a pair of Lam\'e
equations \cite{Lame},
\bea
\ddot{\psi}+(1-2k^2\sn^2t)\psi&=&0 \label{CL} \\
\ddot{\psi}+(1-k^2\sn^2 t)\psi&=&0 \label{OCL}.
\eea
A standard technique when seeking $2K$ and $4K$-periodic solutions of a
Lam\'e equation is to make the following elliptic change of time coordinate,
\beq
t\mapsto x(t)=\am(t)
\eeq
where $\am$ is Jacobi's amplitude function, that is, the unique continuous
function satisfying $\sin x(t)=\sn\, t$, $x(0)=0$. Under this coordinate change
the pair of Lam\'e equations above becomes a pair of Ince equations
\cite{Lame},
\bea
(1+a\cos 2x)\psi''-a\sin 2x\psi'+(1-a+2a\cos 2x)\psi&=&0 \label{CI}\\
(1+a\cos 2x)\psi''-a\sin 2x\psi'+(1+a\cos 2x)\psi&=&0 \label{OCI}
\eea
where $\psi'$ denotes $d\psi/dx$ etc., and $a=k^2/(2-k^2)$ is a period
dependent parameter taking values in $(0,1)$. Note that the coefficients in 
these ODE's are periodic in $x$ with period $\pi$, independent of $k$, and 
hence $T$. So all $k$ dependence is now explicit and resides in the constant
$a(k)$. Odd $2K$ and $4K$-periodic solutions of (\ref{CL},\ref{OCL})
correspond to odd $\pi$ and $2\pi$-periodic solutions of (\ref{CI},\ref{OCI}),
and {\it vice versa.}

So, we 
need to prove that (\ref{CI}) has no odd $2\pi$-periodic solutions, and that
(\ref{OCI}) has neither $2\pi$ nor $\pi$-periodic odd
solutions. Assume, to the
contrary, that such solutions exist. Let $\psi^1$, $\psi^2$ be $2\pi$-periodic
odd solutions of (\ref{CI}) and (\ref{OCI}) respectively, and $\psi^3$ be
a $\pi$-periodic odd solution of (\ref{OCI}). Functions of such parity and
periodicity must have purely sinusoidal Fourier expansions of the following
forms:
\bea
\psi^{\sigma}&=&\sum_{n=1}^{\infty}A_{n}^{\sigma}\sin(2n-1)x\qquad\sigma=1,2
\nonumber \\
\psi^3&=&\sum_{n=1}^{\infty}A_{n}^{3}\sin 2nx.
\eea
Since the coefficients in (\ref{CI}) and (\ref{OCI}) are analytic in $x$,
each $\psi^{\sigma}(x)$ is analytic in a strip of
the complex $x$ plane containing the
real axis \cite{Analytic}. It follows that the Fourier coefficients $A_n^{\sigma}$,
$\sigma=1,2,3$, must satisfy strong decay criteria, namely
\beq
\label{DECAY}
\lim_{n\ra\infty}n^pA_n^{\sigma}=0
\eeq
for every $p\geq 0$. 

Substituting the Fourier expansions for $\psi^{\sigma}$ into their
appropriate Ince equations one obtains the following linear second order
difference equations for $A_n^{\sigma}$:
\bea
A^1_1+A^1_2&=&0,\nonumber\\
A^1_{n+1}+\frac{4n^2-4n+a}{a(2n-1)(n+1)}A^1_n+\frac{n-2}{n-1}A^1_{n-1}&=&0
\qquad n\geq 2; \label{R1} \\
A^2_1+5A^2_2&=&0,\nonumber\\
A^2_{n+1}+\frac{8n(n-1)}{a(4n^2+2n-1)}A^2_n+\frac{4n^2-10n+5}{4n^2+2n-1}
A^2_{n-1}&=&0\qquad n\geq 2;\label{R2} \\
6A^3_1+11aA^3_2&=&0,\nonumber\\
A^3_{n+1}+\frac{2(4n^2-1)}{a(4n^2+6n+1)}A^3_n+\frac{4n^2-6n+1}{4n^2+6n+1}
A^3_{n-1}&=&0\quad n\geq 2\label{R3}.
\eea
\begin{lemma} 
\label{biglemma}
For all $a\in(0,1)$, all nontrivial solutions of the difference
equations (\ref{R1},\ref{R2},\ref{R3}) 
are exponentially divergent. (By trivial we mean
identically zero.)
\end{lemma}
The proof of this lemma is straightforward, but rather involved. The interested
reader is directed to the appendix.

Since none of the sequences $A_{n}^{\sigma}$ satisfy the decay criterion
(\ref{DECAY}), we conclude that no such solutions $\psi^{\sigma}$ exist.
Hence $\K L_0=\K L_{\pm 1}=\{0\}$.

\subsubsection{Invertibility}
\label{sec:invert}
Having established that each $L_n$ as a linear operator $C^-_{T,2}\ra C^-_{T,0}$ is injective, 
we apply standard linear ODE theory to show that each $L_n$ is invertible 
(i.e.\ its inverse is bounded).
Consider $L_n$ acting on the full space $C_{T,2}\ra C_{T,0}$. Each $L_n$ respects the splittings
$C_{T,*}=C^+_{T,*}\oplus C^-_{T,*}$, where $C^+_{T,*}$ denotes the subspace of even loops.
\begin{lemma}
  $\K{L_n} \subset C^+_{T,2}$.
\end{lemma}

\noindent {\it Proof:}
  This is immediate from previous work except for the cases $|n|=1$. Suppose 
$\psi=\psi^++\psi^-\in C_{T,2}$ and 
$L_n\psi=0$. Then since $L_n$ preserves the odd-even splitting, we must have
$L_n\psi^+=L_n\psi^-=0$ and hence $\psi^-=0$, by the previous section.
$\Box$

Let $\K{L_n}^{\perp}= \{\psi\in C_{T,0} : \langle\psi,\K{L_n}\rangle_2=0\}$, where 
$\langle\ ,\ \rangle_2$ is the standard $L^2$ inner product.
\begin{cor}
  $C^-_{T,0}\subset \K{L_n}^{\perp}$.
\end{cor}
\begin{prop}
Each $L_n:C^-_{T,2}\ra C^-_{T,0}$ is an invertible linear map.
\end{prop}

\noindent {\it Proof:}
  Standard linear ODE theory implies that $L_n:C_{T,2}\ra \K{L_n}^{\perp}$ is surjective. 
In particular, $L_n:C_{T,2}\ra C^-_{T,0}\subset\K{L_n}^{\perp}$ is surjective. But since
$L_n$ preserves the odd-even splitting, $L_n:C^-_{T,2}\ra C^-_{T,0}$ is still surjective. 
Now we have a bounded, injective, surjective linear map and hence the closed graph theorem
implies that $L_n$ has a bounded inverse. $\Box$

\subsection{Exponential localization}
\label{sec:local}
We now turn to the question of localization. Issues related to exponential localization in the general class of \emph{networks}
have been treated by Baesens and MacKay in \cite{mackay-baesens}. Localization of our continued periodic solutions will follow 
from an application of their results. To state the relevant 
results from \cite{mackay-baesens} we need to recall some definitions from that paper.

Consider a \emph{network}, that is a countable metric space $S$ with metric $d$ (in our case $S=\mathbb{Z}$ with $d(m,n)=|m-n|$).
To each $s\in S$ associate two normed spaces $X_s$ and $Y_s$, with norms $|.|_{X_s}$ and $|.|_{Y_s}$, to be considered as a local state
space and a local force space respectively -- in our case each $X_s$ is the space of $C^2$ odd $T$-periodic loops in $\R$ with 
the norm given by equation (\ref{eq:c2norm}) and each $Y_s$ is the space of $C^0$ odd $T$-periodic loops in $\R$ with 
the norm given by equation (\eqref{eq:c0norm}).
Define $X$ by 
\begin{equation}
  X = \left\{x\in \oplus_{s}X_s : \|x\|_X := \sup_{s\in S}{|x_s|_{X_s} < \infty }\right\}
\end{equation}
and $Y$ similarly (in our case $X=\ctwoloops$, $Y=\czeroloops$ and the norms on $X$ and $Y$ are just those already given on 
these spaces of loops in section 3). 
\begin{defn}
A bounded linear map $L:X\ra Y$ is said to be \emph{$\phi$-exponentially local} if there exists $\epsilon>0$ and a 
continuous function $\phi:[0,\epsilon)\ra\R$ such that
\begin{equation}
  \sup_{r\in S}{\sum_{s\in S}{|L_{rs}|z^{d(r,s)} \le \phi(z)}}, \quad z\in [0,\epsilon).
\end{equation}
\end{defn}
\noindent
The tridiagonal linear map $DF_{\Theta}:\ctwoloops\ra\czeroloops$ is an exponentially local map 
(e.g. we can take $\phi(z)=\|DF_{\Theta}\|(2z+1)$).
\begin{defn}
  $u\in Y$ (or in $X$) is \emph{$(C,\lambda)$-exponentially localized} about a site $o\in S$, if there exist $C>0, \ 0<\lambda <1$
such that $|u_s|\le C\lambda^{d(s,o)}$.
\end{defn}
\noindent
Certainly, a one-site breather is exponentially localized about the single oscillating site $o$ 
(e.g. we can take $C=|u_o|$, and any $\lambda>0$). We now state the main result of \cite{mackay-baesens}.
\begin{thm}
\label{thm:bm}
(Baesens-MacKay) If $L: X\ra Y$ is $\phi$-exponentially local and invertible, and $u\in Y$ is $(C,\lambda)$-exponentially
localized about site $o\in S$, then $L^{-1}u$ is $(WC,\mu)$-exponentially localized about $o$, where $W$ and $\mu$ depend 
only on $\|L^{-1}\|$, $\lambda$ and a function $\beta$ related to the function $\phi$.
\end{thm}

As an application of the above theorem, MacKay and Baesens prove a theorem 
concerning continuation problems. Let $F:\R \oplus X\ra Y$ be a $C^1$ function, which at 
$\alpha=0$ has a regular zero, i.e.\
 there is a solution $\Theta_0$ of $F(0,\Theta)=0$ at which
$DF_0$ is invertible. Then, by the implicit function theorem there is a neighbourhood of 0 in
which $F(\alpha,\Theta)=0$ has a unique solution $\Theta_{\alpha}$ close to $\Theta_0$. The solution
satisfies
\begin{equation}
  \label{ift}
  \frac{d\Theta_{\alpha}}{d\alpha} = -L^{-1}_{\alpha}\left(\frac{dF}{d\alpha}\left(\Theta_{\alpha}, \alpha\right)\right)
\end{equation}
where
$L_{\alpha}= DF_{\alpha}(\Theta_{\alpha})$
and may be continued as long as $L_{\alpha}$ remains invertible. In this setting they prove:
\begin{thm}
\label{thm:exp}
(Baesens-MacKay)
  Suppose $\Theta_0$ is exponentially localized about site $o\in S$, with continuation $\Theta_{\alpha}$
for $\alpha\in [0,\alpha_0]$ and with $L_{\alpha}$ $\phi$-exponentially local. If 
$\frac{dF}{d\alpha}(\Theta_0)$ is $(A,\mu)$-exponentially localized about $o$, $F\in C^2$, and 
$\frac{d}{d\alpha}DF$ is $\psi$-exponentially local for 
$|\Theta|\le \sup_{\alpha \in [0,\alpha_0]}{|\Theta_{\alpha}|}$ and $\alpha \in [0,\alpha_0]$, then
$\Theta_{\alpha}$ is exponentially localized about $o$ for $\alpha \in [0,\alpha_0]$.
\end{thm}

We have shown that for the function $F$ defined by (\ref{eq:f}), for each  
$T\in(2\pi,\infty)\setminus2\pi\mathbb{N}$ the odd one-site breather, $\bTheta_0$ of period $T$ is
a regular zero of $F$. Choose $\alpha_0>0$ so that for $\alpha\in [0,\alpha_0]$ the 
continuation $\bTheta_{\alpha}$ exists. We claim that the conditions of Theorem \ref{thm:exp}
obtain.

For each $\alpha \in [0,\alpha_0]$, $L_{\alpha}$ is $\phi$-exponentially local where
$\phi(z) = A(2z+1)$ and $A=\max_{\alpha \in [0,\alpha_0]}{\|L_{\alpha}\|}$. 
From
\beq
  \frac{dF}{d\alpha}(\bTheta_0)_n = \left\{
  \begin{array}{ll}
    2\sin{\theta}\, & \mbox{if $n$ = 0},\\
    -\sin{\theta}\, & \mbox{if $|n|$ = 1},\\
    0\, & \mbox{otherwise}
  \end{array}\right.
\eeq
we see that $\frac{dF}{d\alpha}(\Theta_0)$ is exponentially localized about $n=0$.
It follows from 
\bea
  \frac{d}{d\alpha}DF_{\Psi}(\delta\bPsi)_n &=&
 [(\cos{(\psi_{n+1}-\psi_{n})} + \cos{(\psi_{n-1}-\psi_{n})})\,\delta\psi_n 
\nonumber \\ & &
   +\cos{(\psi_{n+1} + \psi_{n})}\, \delta\psi_{n+1} + \cos{(\psi_{n-1} + \psi_{n})}\,\delta\psi_{n-1}]
\eea
that $\frac{d}{d\alpha}DF$ is $\psi$-exponentially local for $\psi(z)=2(1+z)$. It is straightforward
to check that $F$ is $C^2$. Hence Theorem \ref{thm:exp} may be applied in our situation: the 
continued periodic solutions $\bTheta_{\alpha}$ remain exponentially localized and thus are
discrete breathers as claimed. $\Box$

\subsection{Continuation of even breathers}

We have proved that for any $T\in(2\pi,\infty)
\backslash 2\pi\N$ there exists a 
locally unique continuation $\bTheta_{\alpha}$ through $\Omega_{T,2}^-$ of the
odd one-site breather $\bTheta$. The purpose of this section is to prove that
a similar result holds for continuation of {\em even\/} one-site breathers 
also. Since this is not quite so obvious as it may seem, we will state and
prove this result carefully. Let $\Omega_{T,n}$ be the space of all 
$T$-periodic $C^n$ loops, and $\Omega_{T,n}^+$ be its even subspace, both
equipped with their natural norms (see section \ref{sec:fnspaces}). Translating
$\bTheta$, the odd one-site breather of period $T$, through a time of $T/4$
one clearly obtains an even one-site breather, call it $\wt{\bTheta}$
(so explicitly, $\wt{\bTheta}(t)=\bTheta(t+T/4)$. Note this is a slight
abuse of notation: we are using $\bTheta$ to represent both a point in 
$\Omega_{T,2}^-$ and the value of the corresponding function at a particular
time). If
one similarly translates the family $\bTheta_{\alpha}$, one obtains a 
continuation $\wt{\bTheta}_{\alpha}$ of $\wt{\bTheta}$ through breather 
solutions in $\Omega_{T,2}$. We will prove that, for $\alpha$ small enough,
this continuation actually lies in the even subspace $\Omega_{T,2}^{+}$.

\begin{cor} For each $T\in(2\pi,\infty)\backslash 2\pi\N$ there exists
$\epsilon>0$ such that there is a locally unique continuous family 
$\wt{\bTheta}_{\alpha}\in\Omega_{T,2}^+$ of solutions of the TDSG system at
coupling $\alpha\in[0,\epsilon)$, with $\wt{\bTheta}_0=\wt{\bTheta}$. These
solutions are exponentially localized.
\label{cor:even}
\end{cor}

\noindent{\it Proof:} By Theorem \ref{thm:cont} there exists 
$\epsilon>0$ such that 
for $\alpha\in[0,\epsilon)$ there is a continuous family $\bTheta_{\alpha}\in
\Omega_{T,2}^-$ of odd breather solutions with $\bTheta_{0}=\bTheta$. Choosing
$\epsilon$ sufficiently small, each $\bTheta_{\alpha}$ is unique (i.e.\ is the
only zero of $F$) in the open
ball $B_{\epsilon}(\bTheta)\subset\Omega_{T,2}^-$. Consider the family
$\wt{\bTheta}_{\alpha}$ defined by 
$\wt{\bTheta}_{\alpha}(t)=\bTheta_{\alpha}(t+T/4)$. Clearly 
$\wt{\bTheta}_{\alpha}$ is a continuation of the even 
one-site breather $\wt{\bTheta}$ through exponentially localized breathers in
$\Omega_{T,2}$. It remains to prove that this continuation actually lies in
$\Omega_{T,2}^+$.

For each $\alpha\in(0,\epsilon)$ define another breather solution
$\wh{\bTheta}_{\alpha}$ of the TDSG system by $\wh{\bTheta}_{\alpha}(t)=
-\bTheta_{\alpha}(t+T/2).$ By periodicity and oddness of 
$\bTheta_{\alpha}$ one sees that $\wh{\bTheta}_{\alpha}\in\Omega_{T,2}^-$.
Also, ${\|\wh{\bTheta}_{\alpha}-\bTheta\|}_2\equiv
{\|\bTheta_{\alpha}-\bTheta\|}_2<\epsilon$ since the one-site breather
$\bTheta$ satisfies the identity $\bTheta(t+T/2)\equiv-\bTheta(t).$ But the
continuation $\bTheta_{\alpha}$ is unique in the ball $B_{\epsilon}(\bTheta)
\subset\Omega_{T,2}^-$, so we conclude that $\wh{\bTheta}_{\alpha}=
\bTheta_{\alpha}$. Hence $\dot{\bTheta}_{\alpha}(T/2)=
-\dot{\bTheta}_{\alpha}(0)$, and by time reversal symmetry of the TDSG
system, $\dot{\bTheta}_{\alpha}(T/4)=\z$. But $\dot{\wt{\bTheta}}_{\alpha}(0)
=\dot{\bTheta}_{\alpha}(T/4)=\z$, so $\wt{\bTheta}_{\alpha}$ is even. 

Local uniqueness of the even continuation $\wt{\bTheta}_\alpha$ is 
immediate. For if the continuation were not unique on a small enough 
neighbourhood of $\bTheta$, 
one would be able to construct an alternative odd continuation,
again by time translation, in violation of local uniqueness of
$\bTheta_\alpha$. $\Box$

\section{Continuation of one-site breathers: numerical results}
\label{sec:num}
For each non-resonant period $T$, Corollary \ref{cor:even} yields a
1-parameter family of $T$-periodic even breathers, but fails to give 
detailed information about these solutions.
For example, it gives little information about the domain of existence of the 
continuation, how 
this domain varies with the period $T$, or the profiles of the breathers. We
have obtained some such information through numerical work. 
This section describes in 
detail the numerical 
scheme used and the results obtained.

Before describing the numerical work we note that there is one established
mechanism which puts limits on the extent of the continuation being attempted.
Namely, one expects unique continuation to fail once the breather enters into
resonance with the phonons of the system. Linearizing the equations of motion
about the vacuum and seeking solutions of travelling wave type yields the
phonon dispersion relation
\beq
\label{dispersion}
\omega^2=1+4\alpha\sin^2\frac{k}{2}
\eeq
relating frequency $\omega$ to wavenumber $k$ \cite{ward-speight}. From this
we see that the phonon frequency band at coupling $\alpha$ is
$[1,\sqrt{1+4\alpha}]$. Suppose we fix a frequency 
$\omega\in(n^{-1},(n-1)^{-1})$, $n=2,3,\ldots$, 
so that its lowest harmonic above
the bottom edge of the phonon band is $n\omega$. Then during continuation of
the frequency $\omega$ one-site breather, $\alpha$ increases from $0$ and
the phonon band expands, eventually capturing the harmonic $n\omega$ when
$n\omega=\sqrt{1+4\alpha}$ (if continuation persists this far). Beyond this 
point one expects not only loss of uniqueness of the continuation, but also a
breakdown in exponential localization \cite{review}, 
since the phonons themselves are not 
localized. Thus the union of the curves
\beq
\label{boundary}
\alpha=\frac{1}{4}(n^2\omega^2-1),\qquad n^{-1}<\omega<(n-1)^{-1},\quad n=2,3,
\ldots
\eeq
with vertical line segments at the resonant frequencies $\omega=n^{-1}$
provides an upper boundary to the possible breather existence
 domain in the $(\omega,
\alpha)$ parameter space (see figure 1). We will find that the  
existence domain is considerably smaller than the region  bounded by this 
curve.
\begin{figure}
\centerline{\epsfysize=3truein
\epsfbox{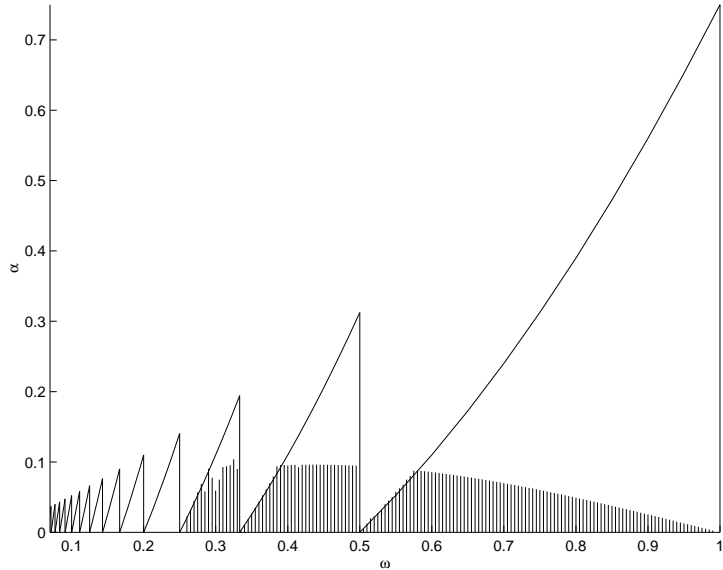}}
\caption{The continuation domain of one-site breathers in $(\omega,\alpha)$
space. The upper, jagged curve is the boundary obtained from the phonon
resonance argument, and consists of parabolic segments. The shaded region
beneath this curve is the actual continuation domain obtained by numerical
analysis for $\omega\in[0.25,1]$, one vertical line for each frequency sampled.
The region to the left of $\omega=0.25$ is inaccessible to our numerical
scheme.}\label{fig:cont}
\end{figure}

The idea behind the numerical method used to map out the existence domain is
to convert the existence proof of section \ref{sec:analytic} into a 
constructive means of finding periodic solutions. By the analysis of that
section, we know that the nonlinear function $F$ defined therein is invertible
at a one-site breather (provided the frequency of the breather is nonresonant)
and hence, by the implicit function theorem, we know that in a neighbourhood
of the one-site breather, $F^{-1}(0)$ is a smooth curve. This curve is the
one-parameter family of periodic solutions we seek, parametrized by the 
coupling $\alpha$. Given one point on the curve (e.g.\ the one-site breather at
$\alpha=0$) one can find sufficiently close neighbouring points by using the
given point as initial input in a root-finding algorithm for $F$. Working
piecemeal from $\alpha=0$, we can in this way construct the entire curve. To
solve $F=0$ we may use a Newton-Raphson method, which will converge provided
$DF$ is invertible at the root and we attempt a small enough step along the
curve.

As it stands, this scheme is impractical since the spaces on which $F$ is 
defined are infinite dimensional. We may seek a finite-dimensional truncation
of the scheme in several ways. One approach is to truncate the lattice to 
finite size, and approximate the space of periodic functions associated with
each lattice site by some finite Fourier series representation. Then the 
problem is reduced to that of determining the finite number of Fourier
coefficients at the finite number of lattice sites, which is solvable by
Newton-Raphson. This is somewhat complicated in the TDSG system because one
must compute Fourier transforms of non-polynomial nonlinear terms, 
necessitating the use of fast Fourier transforms \cite{aubry-aubry}.

For this reason we prefer an alternative approach, 
where $F$ is replaced by a function stemming from the
Poincar\'e return map for period $T$, which we shall denote $\T$. We again
truncate the lattice to finite size (say $2N+1$ sites symmetrically distributed
about the breather centre), and consider the map on phase space which 
maps the initial data of a solution to its phase space position after time
$T$. This map may be computed in practice by solving the lattice equations
of motion using a 4th order Runge-Kutta method. Fixed points of the map
$\T$, or equivalently, zeros of $\T-{\rm Id}$, are clearly periodic solutions.
However, on the whole phase space, solutions are never isolated because of the
time translation symmetry. Consequently, $D\T$ is never invertible, and hence
we cannot directly apply a Newton-Raphson algorithm.

The solution to this difficulty is to consider some restriction of $\T$ which
has no such time-translation symmetry. Marin and Aubry \cite{aubry-marin}
consider the restricted return map $\T_P:\R^{2N+1}\ra\R^{2N+1}$ which takes 
initial positions of time-reversal symmetric solutions (so that initial 
velocity is $0$) to their positions at time $T$. That fixed points of $\T_P$
are truly $T$-periodic is an immediate consequence of energy conservation. An
unfortunate consequence of this choice is that at $\alpha=0$, $D(\T_P-
{\rm Id})$ is singular for all $\omega$. Explicitly,
\beq
D\T_P={\rm diag}(\cos T, \ldots,  \cos T, y_1(T), y_0(T), y_1(T), \cos T, 
\ldots,
\cos T),
\eeq
where $y_0,y_1$ are solutions of (\ref{C},\ref{OC}) respectively with initial
data $y_n(0)=1,\dot{y}_n(0)=0$, $\theta$ now being the {\em even} $T$-periodic
solution of the pendulum equation. Let $\bar{y}_0$ be the odd solution of
(\ref{C}) with $\dot{\bar{y}}_0(0)=1$. Then a short calculation
reveals that $\bar{y}_0(t)=-\theta(t)/\sin\theta(0)$ which is $T$-periodic.
Conservation of the Wronskian $y_0\dot{\bar{y}}_0-\dot{y}_0\bar{y}_0$ then
implies that $y_0(T)=1$. Hence $\det(D\T_P-\IM)=0$, and one is not guaranteed
good behaviour of the Newton-Raphson algorithm for $\T_P-{\rm Id}$ at, and
close to, $\alpha=0$. This problem is not specific to the TDSG system, and
was in fact pointed out in \cite{aubry-marin}.

An alternative restriction, which we denote $\T_V$, arises by mapping the
initial position of a time-reversal symmetric solution to its velocity after
one period. Certainly, $T$-periodic solutions are zeros of $\T_V$, although
the converse is clearly false since the $2T$-periodic one-site breather is a
zero of $T_V$ at $\alpha=0$. (In fact any zero of $T_V$ {\em must} be $2T$-periodic
by time-reversal symmetry of the system.) This is not a problem provided the
zero we seek is isolated, which is guaranteed by the inverse function theorem
so long as $D\T_V$ remains invertible along the continuation curve. The 
advantage of this approach is that at $\alpha=0$,
\beq
D\T_V={\rm diag}(-\sin T, \ldots, -\sin T, \dot{y}_1(T), \dot{y}_0(T),
\dot{y}_1(T), -\sin T,\dots,-\sin T),
\eeq
which turns out to be nonsingular except at the resonant periods $T=2n\pi$,
where continuation is not expected anyway,
and also at the ``false resonances'' $T=(2n+1)\pi$. As continuation proceeds,
care must be taken, once a zero of $\det D\T_V$ has been encountered, that
we do not follow a spurious, non-periodic branch of $\T_V^{-1}(0)$. At the
end of the Newton-Raphson algorithm we check separately that the proposed
periodic solution really is a fixed point of $\T_P$ to within numerical
tolerance (note that this involves no extra computational cost since $T_P$ is
an inevitable by-product of the calculation of $\T_V$). So it is possible
that the continuation could fail prematurely at a zero of $D\T_V$ if it is
diverted from the true curve onto a spurious branch. This turns out to be
a problem only around the false resonant frequency $\omega=\frac{2}{7}$ in our
work (see discussion below).

To test whether the continuation has failed  we use two criteria, beyond mere
failure of the Newton-Raphson algorithm to converge. First, we check that each
new solution does not differ greatly from the previous one, by computing 
$||\bTheta_{\alpha_1}-\bTheta_{\alpha_0}||_2$. This is to ensure that the
continuation has not jumped across to some distant zero of $\T_V$, as
Newton-Raphson output is wont to do close to singular points. Second, we seek
bifurcations of the continuation by applying a simple, necessary, but not
sufficient, test. Namely, we terminate continuation if ever
$\det D\T_V$ and $\det D\T_P$ {\em simultaneously} vanish (note that this test
requires minimal extra computational effort since $D\T_V$ is required by the
Newton-Raphson algorithm, and cannot be calculated without producing $D\T_P$ as
a by-product. The only extra cost is that of the $LU$ decomposition of
$D\T_P$, insignificant compared to the cost
 of the Runge-Kutta method). That this 
test is not sufficient is clear from consideration of the false resonant
frequencies $\omega=\frac{2}{n}$ at $\alpha=0$, where no bifurcation can exist
by Corollary \ref{cor:even}. A full bifurcation analysis, along the lines
of \cite{aubry-aubry}, would remedy this
insufficiency, but only at considerable computational cost, and would be well
beyond the scope of this paper. We hope to perform such an analysis in the
future.

The results of the continuation algorithm, using a 13-site lattice with
fixed endpoints and the implementations of the Runge-Kutta and
Newton-Raphson methods outlined in \cite{recipes},
 are presented in figure 1. The
phonon resonance boundary clearly curtails continuation just to the right
of the resonant frequencies $\omega=\frac{1}{2},\frac{1}{3},\frac{1}{4}$.
However for large regions of $\omega$ space, continuation fails before phonon
resonance can occur, at what appear to be genuine bifurcations (at least
 $\det D\T_V=\det D\T_P=0$) the underlying physical mechanism for
which remains mysterious. The exception is a small $\omega$ range around the
false resonance $\omega=\frac{2}{7}$ where the continuation seems to fail 
spuriously
due to its encountering a zero of $\det D\T_V$ which is not a genuine
bifurcation point ($\det D\T_P\neq 0$), but which derails the Newton-Raphson
algorithm nonetheless. The $\omega\in[0.27,0.3]$ 
part of figure 1 should thus be 
viewed with a little scepticism.
\begin{figure}
\centerline{\epsfysize=2truein
\epsfbox[55   404   598   600]{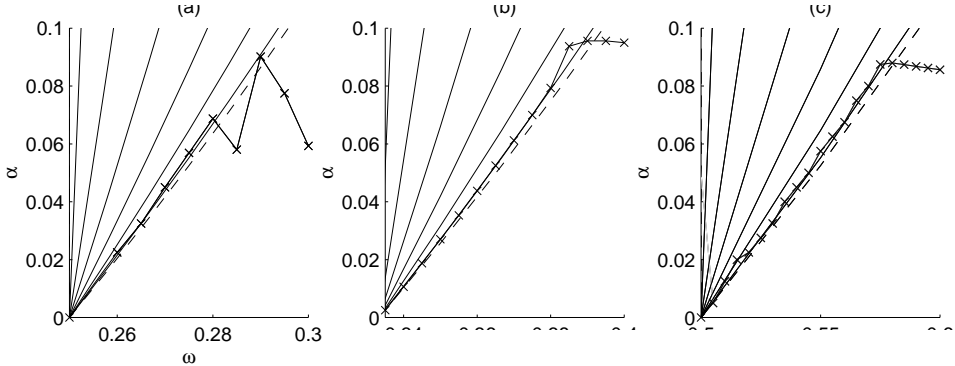}}
\caption{Details of the continuation domain around three resonant frequencies:
(a) $\omega=1/4$, (b) $\omega=1/3$ and (c) $\omega=1/2$. In each case, 
the solid curve with
crosses represents the numerical results, the dashed curve is the phonon
resonance boundary for an infinite lattice, and the six solid curves without
crosses are the parabolic resonance curves for the 13-site lattice.}
\label{fig:detail}
\end{figure}

Closer examination of the results around 
$\omega=\frac{1}{2},\frac{1}{3},\frac{1}{4}$ (figure 2) reveals 
that the continuation slightly oversteps the resonance boundary defined by
equation (\ref{boundary}). This can be well understood as a finite size effect:
equation (\ref{dispersion}) is the dispersion relation for an infinite 
lattice, whereas we have a $2N+1$ site lattice, divided by the central site
into two ``tails'' of length $N$ which remain close to the vacuum. The relevant
spectrum for small amplitude oscillations
is doubly degenerate, consisting of two copies of the 
normal mode spectrum of the
$N$-site linearized lattice (one copy for each of the two tails), 
which itself consists of $N$ distinct frequencies
$\omega_j$, $j=1,2,\ldots,N$, easily shown \cite{casimir} to satisfy
\beq
\omega_j^2=1+4\alpha\sin^2\left(\frac{j\pi}{2N+2}\right).
\eeq
So bifurcation should be expected at that value of $\alpha$ for which
$\omega_N$ coincides with the lowest harmonic of $\omega$ above $1$. This 
leads to a slightly different boundary from that previously discussed. Since
the ``phonon'' spectrum is now discrete, it is possible for the continuation to
skip over this first bifurcation point occasionally, and detect a later one
instead, due to $\omega_{N-1}$, $\omega_{N-2}$ etc. In our case, $N=6$, so 
there are six parabolic resonance curves, one for each normal mode
frequency. These are included in figure 2.

As a by-product of the numerical work described above
 we also obtain  breather initial profiles within the continuation domain.
The resulting profiles are highly spatially localized, and
remarkably uniform, all exhibiting basically the same shape: compared to the one-site breather 
of the same period, the central site of the continued breathers is displaced slightly more than in the one-site  
breather, the two adjacent sites are displaced negatively by a small amount, and the remaining sites are barely displaced at all.
Figure 3 shows breather profiles of the same frequency at increasing coupling, and 
profiles at the same coupling and various frequencies. Breather
shape will be discussed further in the next section.
\begin{figure}
\centerline{\epsfysize=2truein
\epsfbox[63   404   548   600]{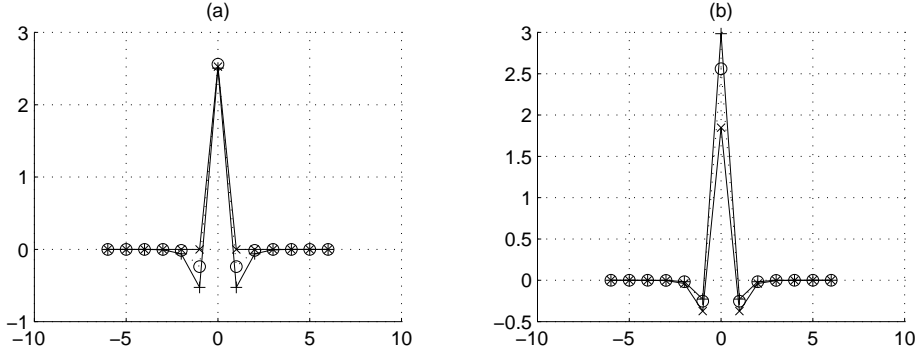}}
\caption{Initial profiles of  breathers, (a) at frequency 0.6 with
 coupling $h=2.0$ (x),
$h=1.8$ (o) and
$h=1.73$ (+); and (b) at coupling 
$h=1.73$ with frequency
$\omega=0.8$ (x),
$\omega=0.52$ (o) and
$\omega=0.44$ (+).}\label{fig:constfreq}
\end{figure}

Finally, we remark that the numerical methods described above are inefficient 
for investigating the long period breathers of the system. 
Indeed, if one used the same time step in the numerical integration schemes 
independent of the period, $T$, then of course the 
time to compute $\T_V$ and $D\T_V$, which occupies the vast majority of the 
computing time,
would grow linearly with $T$. If one wanted to investigate long period 
breathers then it would seem 
more efficient to switch from the Runge-Kutta scheme to a 
symplectic integrator,
of which one expects reliable 
long-time integration at modest time step size. The efficiency of the algorithm
could also be improved by enabling it to take larger steps in $\alpha$ so that
the bifurcation edge is reached more quickly. One method to do this 
(resulting in a so-called Euler-Newton scheme) is to
 generate the 
initial guess for the Newton-Rapshon algorithm by extrapolating along the 
tangent vector to the continuation curve at the last established point. 
Calculating this vector involves solving an extra set of coupled linear
ODEs, but it is thought that the benefit of an enlarged parameter
step length generally outweighs this extra computational cost
\cite{continuation}.

\section{The direction of continuation: a linear calculation}

Corollary \ref{cor:even} shows that for each even one-site breather $\bTheta$
of 
nonresonant period $T$
there exists a family $\bTheta_\alpha\in\Omega_{T,2}^+$, for $\alpha$ in a 
neighbourhood of $0$, satisfying the continuation equation
\begin{equation}
\label{7}
F(\bTheta_\alpha,\alpha)=\z,
\end{equation}
with $\bTheta_0=\bTheta$.
In this section we describe a simple method for computing the {\em direction}
of continuation away from $\bTheta$, 
that is $d\bTheta_\alpha/d\alpha|_{\alpha=0}
\in\Omega_{T,2}^+$, 
which will henceforth be denoted $\bTheta'=(\chi_{n})_{n\in\Z}$.
The method has previously been used to provide the ``predictor'' step in
predictor-corrector schemes for continuation problems 
mentioned above \cite{continuation},
although we shall require only a greatly simplified version since we consider
only the case $\alpha=0$.
Differentiating equation (\ref{7}) with respect to $\alpha$ at $\alpha=0$
yields
\begin{equation}
\frac{\cd F}{\cd\alpha}(\bTheta,0)+DF_{\Theta}\bTheta'=0.
\end{equation}
Denoting the $T$-periodic even solution of the pendulum equation $\theta$, one
easily obtains the following explicit expressions,
\begin{equation}
\left[\frac{\cd F}{\cd\alpha}(\bTheta,0)\right]_{n}=\left\{
\begin{array}{ll}
2\sin\theta & n=0 \\
-\sin\theta & n=\pm 1 \\
0 & |n|>1,
\end{array}\right.
\end{equation}
and
\begin{equation}
\left[DF_\Theta\bTheta'\right]_{n}=\left\{
\begin{array}{ll}
\ddot{\chi}_{n}+\cos\theta\chi_{n} & n=0 \\
\ddot{\chi}_{n}+\frac{1}{2}(1+\cos\theta)\chi_{n} & n=\pm 1 \\
\ddot{\chi}_{n}+\chi_{n} & |n|>1.
\end{array}\right.
\end{equation}
So inverting $DF_\Theta$ to find $\bTheta'$ 
amounts to solving an infinite set of
ordinary differential equations,
\bea
\label{9}
\ddot{\chi}_{0}+\cos\theta\chi_{0}&=&-2\sin\theta \\
\label{10}
\ddot{\chi}_{\pm 1}+\frac{1}{2}(1+\cos\theta)\chi_{\pm 1}&=&\sin\theta \\
\label{11}
\ddot{\chi}_{n}+\chi_{n}&=& 0\qquad |n|>1,
\eea
in the space $\Omega_{T,2}^+$. Provided $T\notin 2\pi\Z$, equation (\ref{11})
has only the trivial solution $\chi_{n}=0$, so to first order the continuation
leaves the $|n|>1$ sites fixed at $0$.

For the $|n|\leq 1$ equations (\ref{9},\ref{10}), the question remains how one
should choose initial data $\chi_{n}(0)\in\R$, $\dot{\chi}_{n}(0)=0$ in order
to generate a $T$-periodic solution. Let $y_{n}(t)$ be the solution of the
homogeneous problem with initial data $y_{n}(0)=1$, $\dot{y}_{n}(0)=0$,
and $Y_{n}(t)$ be the particular integral of the inhomogeneous equation
with initial data $Y_{n}(0)=\dot{Y}_{n}(0)=0$. Then
\begin{equation}
\chi_{n}(t)=\chi_{n}(0)y_{n}(t)+Y_{n}(t)
\end{equation}
and demanding that $\chi_{n}$ be periodic gives an overdetermined pair of
linear equations,
\bea
\label{13}
\chi_{n}(0)&=&y_{n}(T)\chi_{n}(0)+Y_{n}(T) \\
\label{14}
0&=&\dot{y}_{n}(T)\chi_{n}(0)+\dot{Y}_{n}(T)
\eea
for $\chi_{n}(0)$ in terms of the constants $y_{n}(T),\dot{y}_{n}(T),Y_{n}(T),
\dot{Y}_{n}(T)$. Equation (\ref{14}) can be solved for $\chi_{n}(0)$, leaving
(\ref{13}) as a linear consistency condition for the constants. Clearly 
$\chi_{-1}(0)=\chi_{1}(0)$.

So the direction of change of the initial profile of a breather can be deduced
by solving a coupled set of five second order ODE's (the pendulum
equation, and the homogeneous and inhomogeneous equations for $n=0$ and $n=1$),
over the time interval $[0,T]$, computing the end-point constants and solving
(\ref{14}). This has been done numerically using the fourth order Runge-Kutta
scheme implemented by Maple. The results are presented graphically in
figure 4, a plot of $\chi_{0}(0)$ and $\chi_{\pm 1}(0)$ against frequency
$\omega=2\pi/T$. Note that for all $\omega$, $\chi_{0}>0$ and $\chi_{\pm 1}
<0$, so the continuation starts by pulling the two off-central sites
downwards, and the central site further upwards, as previously observed.
Both $\chi_{0}$ and $\chi_{\pm 1}$ appear to grow unbounded as $\omega\ra 1$
(the small amplitude limit) in agreement with the observation that the length
of the continuation neighbourhood vanishes in this limit. The graph suggests
that as $\omega\ra 0$ ($T\ra\infty$), $\chi_{0}$ vanishes, while $\chi_{\pm 1}$
tends to a nonzero constant. Of course, this limit is numerically inaccessible,
so these remarks are necessarily speculative. One can estimate $\chi_{n}(0)$
from the numerical work of section \ref{sec:num} by comparing $\bTheta$ with
$\bTheta_\alpha$ after the first continuation step. The results of this 
calculation are also presented in figure 4 (broken curve) for purposes of
comparison with the linear method. 
\begin{figure}
\centerline{\epsfysize=3truein
\epsfbox{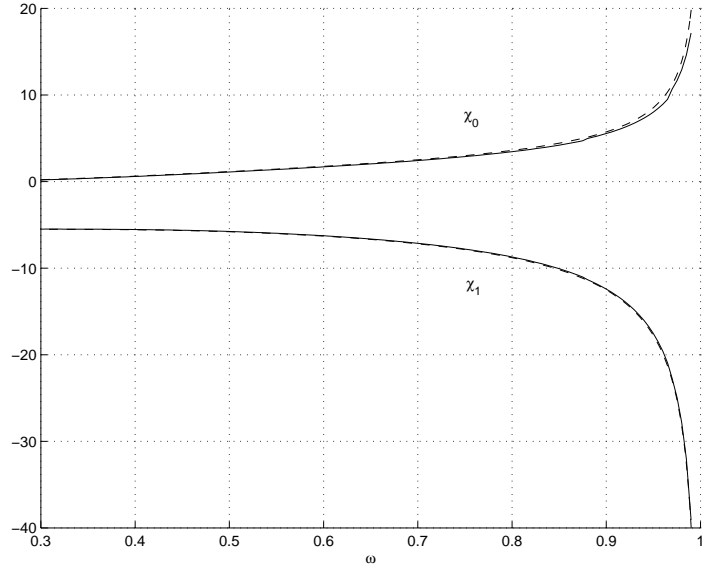}}
\caption{The components $\chi_0$ and $\chi_1$ of the tangent vector 
$\bTheta'$ as functions of frequency, $\omega$. The solid curves were
generated using the linear method, the dashed curves from the breather
profiles generated by the numerical scheme described in section
\ref{sec:num}. The two data sets agree closely (especially for $\chi_1$).}
\label{fig:linear}
\end{figure}
 
\section{Conclusion}

In this paper we have adapted the methods of MacKay and Aubry to prove 
existence of breathers in the TDSG system, despite the fact that there is no
limit in which the system decouples into independent oscillators (so the
existence theorems so far formulated \cite{aubry-mackay,mackay-sepulchre} do
not directly apply). Numerical analysis has revealed that 
the persistence of these breathers away from the $h=2$
limit depends critically on their period. One reason for this is that 
the coupling at which the
breather enters into resonance with the phonons depends on its frequency. 
However, for
most frequencies the continuation actually fails before
it reaches the phonon resonance region. The underlying mechanism for these
early bifurcations is at present unknown. A full bifurcation
analysis may prove informative.

In no case does existence persist for all
$h\in(0,2]$, so the breathers found here are unconnected with
the breathers of the continuum sine-Gordon system (which the TDSG system
approaches as $h\ra 0$). Indeed, they look qualitatively different from 
continuum breathers in that they have sombrero rather than hump shaped initial
profiles. In this respect they appear to differ also from breathers found
in previous studies of spatially discrete systems. By means of a simple
linear method, we have confirmed the uniform trend towards sombrero-shaped
breathers in the TDSG system. It would be interesting to apply the same linear
method to a variety of more conventional oscillator networks, since this 
allows one to examine the direction of continuation of one-site breathers
away from the weak coupling limit with a minimum of computational effort.

\vspace{0.75cm}
\noindent
{\bf Acknowledgments:} The authors would like to thank Rafael de la Llave
for useful discussions, and the referee for suggesting substantial improvements
to an earlier draft of section 4. 
MH acknowledges financial support in the form of a
research studentship from the States Education Authority of Guernsey, and
a University Continuing Fellowship from the University of Texas at
Austin.

\vspace{1cm}
\noindent
\Large
{\bf Appendix}

\normalsize
\vspace{0.75cm}
\noindent
The proof of invertibility of $DF_{\Theta}$ rested on the following lemma,
whose proof we now give.

\vspace{0.25cm}
\noindent
{\bf Lemma \ref{biglemma}}
{\it For all $a\in(0,1)$, all nontrivial solutions of the
linear difference equations
\begin{enumerate}
\item[{\rm (a)}]$A^1_1+A^1_2=0,\quad
\ds{A^1_{n+1}+\frac{4n^2-4n+a}{a(2n-1)(n+1)}A^1_n+\frac{n-2}{n-1}A^1_{n-1}=0
\qquad n\geq 2,}$
\item[{\rm (b)}]
$
A^2_1+5A^2_2=0,\quad \ds{
A^2_{n+1}+\frac{8n(n-1)}{a(4n^2+2n-1)}A^2_n+\frac{4n^2-10n+5}{4n^2+2n-1}
A^2_{n-1}=0\qquad n\geq 2,}$
\item[{\rm (c)}]
$
6A^3_1+11aA^3_2=0,\quad
\ds{
A^3_{n+1}+\frac{2(4n^2-1)}{a(4n^2+6n+1)}A^3_n+\frac{4n^2-6n+1}{4n^2+6n+1}
A^3_{n-1}=0\quad n\geq 2,}$
\end{enumerate}
are exponentially divergent.}

\vspace{0.25cm}
\noindent
{\it Proof:} Each of the difference equations (a), (b), (c) is of a type
subject to the following theorem of Poincar\'e \cite{Poi}:
%
\begin{thm}\, {\it (Poincar\'e)\,
Let $f_{1},f_{2}$ be functions with limits $p_{1},p_{2}$ respectively as
$n\ra\infty$, such that the polynomial $P(\lambda)=\lambda^2+p_1\lambda+p_2$
has two distinct real roots $\lambda_1,\lambda_2$. Then
 any solution of the linear second order
difference equation
$y_{n+1}+f_1(n)y_n+f_2(n)y_{n-1}=0$
has the asymptotic behaviour
$\lim_{n\ra\infty}\frac{y_{n+1}}{y_{n}}=\lambda_i$
for either $i=1$ or $i=2$.}
\end{thm}

In each of the three cases of interest here, $P$ is the same polynomial,
namely $P(\lambda)=\lambda^2+\frac{2}{a}\lambda+1$. For $a\in(0,1)$, this
has two distinct negative roots whose product is unity. Let $\lambda_1$ 
denote the larger root, so that $|\lambda_1|<1$ while $|\lambda_2|>1$.
To prove exponential divergence of $A^{\sigma}$, $\sigma=1,2,3,$
 it is convenient to define
the quotient sequences
\beq
B^{\sigma}_n=-\frac{A^{\sigma}_{n+1}}{A^{\sigma}_n},
\eeq which satisfy first order nonlinear recurrence relations. Note that
the sequence $B^{\sigma}$ exists if and only if it never vanishes. We
will assume that these sequences exist, and justify the assumption
{\it a posteriori}. By Poincar\'e's
Theorem, for each $\sigma$ the sequence $B^{\sigma}\ra|\lambda_i|$ for
either $i=1$ or $i=2$. To prove exponential divergence of $A^{\sigma}$
therefore, it suffices to show that $B^{\sigma}$ cannot converge to a limit
less than 1.

\vspace{0.2cm}{\bf Part (a):} From the difference equation (a), one finds that
$B^1$ satisfies the first order recurrence relation
\beq
B^1_n=\frac{4n^2-4n+a}{a(2n-1)(n+1)}-\frac{n-2}{n+1}\frac{1}{B^1_{n-1}}
\eeq
with initial condition $B^1_1=1$. We claim that $B^1_n\geq 1$ for all $n\in\N$.
The proof is by induction. Assume $B^1_{n-1}\geq 1$. Then
\bea
B^1_n&=&\frac{4n^2-4n}{a(2n-1)(n+1)}+\frac{1}{(2n-1)(n+1)}-\frac{n-2}{n+1}
\frac{1}{B^1_{n-1}} \nonumber \\
&>&\frac{4n^2-4n+1}{(2n-1)(n+1)}-\frac{n-2}{n+1}=1.
\eea
Since $B^1_n\geq 1$ for all $n\in\N$, the sequence exists (never vanishes)
and cannot converge to $|\lambda_1|<1$. Hence $B^1\ra|\lambda_2|>1$, and
$A^1$ is exponentially divergent.

\vspace{0.2cm}{\bf Part (b):} From the difference equation (b), one finds that
$B^2$ satisfies the first order recurrence relation
\beq
B^2_n=\frac{8}{a}\frac{n(n-1)}{4n^2+2n-1}-\frac{4n^2-10n+5}{4n^2+2n-1}
\frac{1}{B^2_{n-1}}
\label{B2}
\eeq
with initial condition $B^2_1=\frac{1}{5}$. Since the solution of (\ref{B2})
depends on the parameter $a$ we will denote it $B^2(a)$, and further, allow
$a$ to take all values in $(0,1]$, denoting $B^2(1)$ by $\B$. The first
step is to prove that $\B$ exists (is nonvanishing) and does not converge
to a limit less than $1$. Assume $\B_{n-1}>n/(n+1)$. Then for all $n\geq 2$
\bea
\B_n-\frac{n+1}{n+2}&>&\frac{8n(n-1)}{4n^2+2n-1}-\frac{4n^2-10n+5}{4n^2+2n-1}
\frac{n+1}{n} \nonumber \\
&=&\frac{6n-10}{(4n^2+2n-1)(n+2)n}>0.
\eea
Explicit evaluation of the first few terms yields that
$\B_3=\frac{37}{41}>\frac{4}{5}$. Hence, by induction, for all $n\geq 3,$
$\B_n>(n+1)/(n+2)$, so $\B$ exists and does not converge to a limit less
than $1$. 

Now consider the sequence $B^2(a)$ for $a<1$. We claim that each term
$B^2_n(a)$ is a nonincreasing function on $(0,1]$. To see this, observe that
$B^2_1(a)=\frac{1}{5}$ is manifestly nonincreasing, and assume that
$B^2_{n-1}(a)$ is nonincreasing for some $n\geq 2$. Then $B^2_{n-1}(a)\geq
B^2_{n-1}(1)=\B_{n-1}
>0$ by the paragraph above, so $B^2_{n-1}(a)$ is positive and
nonincreasing. From the relation (\ref{B2}) it immediately follows that
$B^2_{n}(a)$ is nonincreasing, so the claim is proved. Hence $B^2(a)$ is
strictly positive (and hence exists) for all $a\in(0,1)$. Further
$\lim B^2(a)\geq\lim\B\geq 1>|\lambda_{1}|$, so $B^2\ra|\lambda_2|$ and
$A^2$ is exponentially divergent. 

\vspace{0.2cm}{\bf Part (c):} The argument is almost identical to part (b).
The sequence $B^3(a)$ satisfies
\beq
B^3_n=\frac{2}{a}\frac{4n^2-1}{4n^2+6n+1}-\frac{4n^2-6n+1}{4n^2+6n+1}
\frac{1}{B^3_{n-1}}
\label{B3}
\eeq
with initial condition $B^3_1(a)=\frac{6}{11a}$. Once again we allow $a$ to
take all values in $(0,1]$ and denote $B^3(1)$ by $\B$. We claim that for
all $n\in\N$, $\B_n>n/(n+1)$. Clearly, the claim holds true for $n=1$. 
Assume $\B_{n-1}>(n-1)/n$ for some $n\geq 2$. Then
\bea
\B_n&>&2\frac{4n^2-1}{4n^2+6n+1}-\frac{4n^2-6n+1}{4n^2+6n+1}\frac{n}{n-1}
\nonumber \\
\Rightarrow
\B_n-\frac{n}{n+1}&>&\frac{2}{(4n^2+6n+1)(n+1)(n-1)}>0,
\eea
so that $\B_n>n/(n+1)$ and the claim is proved. Hence $\B$ is strictly positive
(hence exists) and cannot converge to a limit less than $1$.

We next claim that for each $n\in\N$, $B^3_n(a)$ is a decreasing function
on $(0,1]$. Clearly $B^3_1(a)=\frac{6}{11a}$ is decreasing. Assume that
$B^3_{n-1}(a)$ is decreasing. Then $B^3_{n-1}(a)\geq B^3_{n-1}(1)=\B_{n-1}>0$,
so $B^3_{n-1}$ is a positive decreasing function of $a$. It follows from
(\ref{B3}) that $B^3_n(a)$ is also decreasing, and the claim is proved. 
Hence $B^3(a)$ is
strictly positive (and hence exists) for all $a\in(0,1)$. Further
$\lim B^3(a)\geq\lim\B\geq 1>|\lambda_{1}|$, so $B^3\ra|\lambda_2|$ and
$A^3$ is exponentially divergent. $\Box$


\begin{thebibliography}{xx}
\bibitem{hill}
                  G.W. Hill,
                  ``On the Part of the Motion of the Lunar
                  Perigee which is a Function of the Mean Motions of
                  the Sun and the Moon'',
                  {\sl Acta Math.} {\bf 8} (1886) 1-36.
\end{thebibliography}
\end{document}